\documentclass[twoside,fleqn]{article}
\usepackage{espcrc2}
\usepackage{graphicx}
\usepackage{amsmath,amssymb}

\title{The hidden strangeness mechanism in 
$D_s^+ \to \omega \pi^+$ and $D_s^+ \to \rho^0 \pi^+$ decays}

\author{
S.\ Fajfer\address[FMF]{Physics Department, University of Ljubljana,
SI-1001 Ljubljana, Slovenia}\address[IJS]{J.\ Stefan Institute,
SI-1000 Ljubljana, Slovenia}\thanks{Talk given by S.\ Fajfer},
A.\ Prapotnik\addressmark[IJS],
P.\ Singer\address[TEC]{Department of Physics, Technion -- Israel
Institute  of Technology, Haifa IL-32000, Israel}, and
J.\ Zupan\addressmark[IJS],\addressmark[TEC]}

\begin{document}

\begin{abstract}

We study possible contributions to the  
$D_s^+ \to \omega \pi^+$ and $D_s^+ \to \rho^0 \pi^+$ 
decay amplitudes. 
The $D_s^+ \to \omega \pi^+$ decay amplitude vanishes when  
naive   factorization is used, 
while the $D_s^+\to \rho^0 \pi^+$ 
decay amplitude arises due to  the annihilation contribution. 
 We find  that  amplitudes for both decays 
might be a result of the  internal  $K$, $K^*$  exchange. 
The $D_s^+ \to \omega \pi^+$  amplitude might obtain 
additional contributions  from 
$D_s^+ \to \rho^+ \eta (\eta')$ re-scattering. The low 
experimental bound on the 
 $D_s^+\to \rho^0 \pi^+$ rate can be understood as a result of 
 combination of
 the $\pi(1300)$ pole dominated annihilation contribution 
  and the $K,K^*$ internal exchanges.  
 The calculated  branching fractions for $D_s^+ \to \omega \pi^+$ and  $D_s^+\to \rho^0 \pi^+$
are in agreement with the current experimental results. 

\end{abstract}

\maketitle

The weak nonleptonic decays of charm mesons were usually approached 
within the  factorization ansatz 
\cite{bsw,buccel1a1,buccella2,cheng,hin,rosner}. 
A decade ago it was realized \cite{buccel1a1,buccella2} 
that one has to include the 
 effects of 
final state interactions (FSI), with the simplest approach being to 
treat  the 
FSI by assuming the dominance of nearby resonances. 
This leads to  rather good overall agreement with
the experimental data \cite{buccel1a1,buccella2};  
however, there are a few cases where none of the existing  
approaches work. Two such examples are the channels (quoting 
the PDG experimental values \cite{PDG})
\begin{equation} 
\begin{split}
 BR(D_s^+ \to \omega \pi^+) &= (2.8 \pm 1.1) \times 10^{-3},
\\
 BR(D_s^+ \to \rho^0 \pi^+) &< 7 \times 10^{-4}.
\label{e1}
\end{split}
\end{equation}

The current  theoretical approaches usually predict  that the 
$D_s^+\to \rho^0 \pi^+$ branching fraction is equal 
\cite{buccella2} or even larger than 
the branching fraction 
 for the $D_s^+ \to \omega \pi^+$ decay \cite{hin,rosner} in
contradiction with the
present data \eqref{e1}. 

On the other hand, the observation of the $D^+_s\to \omega \pi^+$ 
decay (\ref{e1}) has been motivated  as a clean signature of the annihilation decay of  
$D_s^+$ \cite{balest}. The sizes of annihilation contributions are very important for phenomenological 
studies, but  are also very hard to obtain from 
theoretical considerations (see e.g., \cite{sinha}). Understanding the 
origin of the $D^+_s\to
\omega \pi^+$ transition is thus of great theoretical interest.

Let us first discuss the two modes \eqref{e1} using factorization approximation
for the weak vertex. In this approximation the $D_s^+ \to \omega
\pi^+$ amplitude is zero due to G - parity conservation, which gives a
vanishing 
$\langle \omega \pi^+ |({\bar u} d)_{V-A} |0\rangle$  matrix element  \cite{FPPZ}.
The $D_s^+ \to \rho^0 \pi^+$ decay amplitude, on the other hand, already in the
factorization limit
receives a  contribution through the annihilation graph,  Fig. 1,
 \begin{equation} 
\begin{split}
 {\cal M} (D_s^+ \to \rho^0 \pi^+)=
 \frac{G_F}{{\sqrt 2}}& V_{us} V_{ud}^*a_1 \times\\
 \times \langle \rho^0 \pi^+ | 
 ({\bar u} d)_{V-A} |0\rangle &\langle 0| ({\bar s} c)_{V-A}|D_s^+\rangle, 
\label{e3}
\end{split}
\end{equation} 
leading to simple $\pi$ pole dominance
in the \break $\langle \rho^0 \pi^+ |({\bar u} d)_{V-A} |0\rangle$
matrix element.
The analysis of \cite{gourdin} indicates that 
$\pi(1300)$ states dominate this annihilation graph, while the contribution of 
 the lowest lying $\pi$ is negligible. 
In \cite{FPPZ} we have estimated 
the size of  the annihilation 
contribution 
 coming from the $\pi(1300)$ 
intermediate state. We found  $f_{\pi (1300)}< 4$ MeV \cite{FPPZ}. In the 
factorization approximation for the weak vertex 
 we then get 
\begin{equation}
BR(D_s^+ \to \rho^0\pi^+)_{\pi(1300)}< 7 \times 10^{-4}\,, \label{annihl}
\end{equation}
where we have used $f_{D_s}=230$ MeV, together with 
the conservative assumptions of $BR(\pi(1300)\to \rho \pi)\sim 100\%$
and $\Gamma(\pi(1300))$ equal to its upper experimental bound of $600$
MeV. 
The interference with other 
annihilation contributions from intermediate $\pi$ and $\pi(1800)$ states can 
somewhat change the above estimate (using PCAC, the contribution from $\pi$ was 
found  in \cite{gourdin} to be negligible, while the contribution of  
$\pi(1800)$ is difficult to estimate due to the lack of 
experimental data). In addition, also the  FSI contributions (to be
considered shortly) fall in exactly the same range  \cite{FPPZ}.  
Therefore,  unless there are large 
cancellations, the value of  $BR(D_s^+ \to \rho^0\pi^+)$ is expected 
to be near 
to its present experimental upper bound \eqref{e1}.  

\begin{figure}
\begin{center}
\includegraphics[width=6cm]{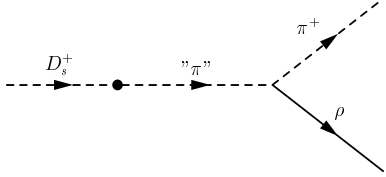}\\[-8mm]
\caption{\footnotesize{Annihilation diagram of 
$D_s \to \rho \pi$ decay.
}}
\end{center}
\end{figure}

In the case of the $(\omega \pi^+)$ final state there is no such  resonance 
annihilation contribution and one has to explain a 
relatively large experimental value for $BR(D_s\to \omega \pi^+)$
\eqref{e1} in
a different  way. An important observation is  that there are 
multi-body intermediate states that do have the correct values of $I^G$ and 
$J^P$, for instance the two-body $K^{(*)}\bar{K}^{(*)}$ states. As
we will show in the rest of the talk, it is  possible to explain
the experimental value for $BR(D_s\to \omega \pi^+)$ by considering
the contributions due to the rescattering of these intermediate states.
  
In estimating the contributions from hidden strangeness intermediate 
states (that can arise from spectator quark diagrams), we 
use the following assumptions 
\begin{itemize} 

\item We consider only contributions coming from 
two body intermediate states with $s, \bar{s}$ quantum 
numbers (lowest lying pseudoscalar and vector states). Note that the re-scattering  
through intermediate $K, K^*$ states is possible for both  
$\rho^0 \pi^+$ as well as $\omega \pi^+$ final state, 
while the re-scattering with intermediate $\eta$ 
or $\eta'$ is possible only in the case of $\omega \pi^+$ final state 
due to isospin and $G$ parity conservation. 
\item For the weak transition 
$D_s^+\to (K^{(*)}\bar{K}^{(*)})^+$ in the $D_s^+\to (K^{(*)}\bar{K}^{(*)})^+\to\rho^0\pi^+$ and $D_s^+\to (K^{(*)}\bar{K}^{(*)})^+\to 
\omega\pi^+$ decay chains as well as for the weak transition $D_s^+\to 
\eta(\eta')\rho^+$ in the  $D_s^+\to \eta(\eta')\rho^+\to 
\omega\pi^+$ decay chain we will use the factorization approximation. The 
weak Lagrangian is therefore 
\begin{equation} 
\begin{split}
{\cal L}_{\rm weak}&=-\frac{G_F}{\sqrt{2}} V_{cs} V_{ud}^*\times\\
&\times \big(a_1 
(\bar{u}d)_H (\bar{s} c)_H +a_2 
(\bar{s}d)_H (\bar{u} c)_H \big)\,, 
\end{split}
\end{equation} 
with $(\bar{u}d)_H, \dots$ the hadronized V-A weak currents, $  V_{ij}$ the CKM matrix elements and $a_{1,2}$ the effective (phenomenological) 
Wilson coefficients taken to be $a_1=1.26$  and $a_2= -0.52$  \cite{bsw,buccel1a1,buccella2}.  
 
\item 
Finally, the strong interactions are taken into account through the 
following effective Lagrangian  \cite{chi1,chi2,chi3}: 
\begin{equation}
\begin{split}
{\cal L}_{\rm strong}=
\frac{ig_{\rho\pi\pi}}{{\sqrt 2}}&
Tr(\rho^\mu[\Pi,\partial_\mu \Pi])
\\
-4\frac{C_{VV\Pi}}{f}&\epsilon^{\mu\nu\alpha\beta} 
Tr(\partial_\mu \rho_\nu \partial_\alpha \rho_\beta \Pi)\;,
\label{strong}
\end{split} 
\end{equation} 
where $\Pi$ and $\rho^\mu$ are $3 \times 3$ matrices 
containing pseudoscalar and vector meson 
operators respectively  and $f$ is a 
pseudoscalar decay constant. We used numerical values $C_{VV \Pi}=0.33$, and $g_{\rho\pi\pi}= 5.9$ 
\cite{chi1,chi2,chi3}. 
 
\end{itemize} 
 
In addition we have checked that the use of factorization 
 for  the 
$D_s^+ \rightarrow K^{*+} \bar K^{*0}$, $D_s^+ \rightarrow K^+ \bar K^{*0}$  
and $D_s^+  \rightarrow \bar K^0 K^{*+}$  decays gives reasonable 
estimates of the measured rates (note that we do not need  
$D_s^+ \rightarrow \bar K^0 K^{+}$ in further considerations) \cite{FPPZ}. In these results the 
annihilation contributions have been neglected since they are an order 
of magnitude smaller. 

The situation in the case of $\eta,\eta'$ intermediate states is not 
so favorable. To treat the $\eta,\eta'$ mixing we use the 
approach of Ref.\cite{kroll1} with the value of the 
 mixing angle transforming between $\eta, \eta'$ and $\eta_q\sim(u\bar{u}+d\bar{d})/\sqrt{2}, \eta_s\sim s\bar{s}$  
states taken to be $\phi=40^\circ$.  The factorization approach then 
 gives a reasonable description of $D_s^+ \to \rho^+ \eta$ decay, 
while it does not reproduce 
 satisfactorily  the experimental result for 
 $D_s^+ \to \rho^+ \eta'$ \cite{FPPZ}. 
 This is a known problem as the $D_s^+ \to \rho^+ \eta'$ rate is  very difficult to 
reproduce in  any of the present approaches \cite{buccel1a1,buccella2,hin}. 
This inevitably introduces some further uncertainty into our approach, 
yet the resulting uncertainty is not  
expected to affect significantly our main conclusions.

\begin{figure} 
\begin{center} 
\includegraphics[width=8cm]{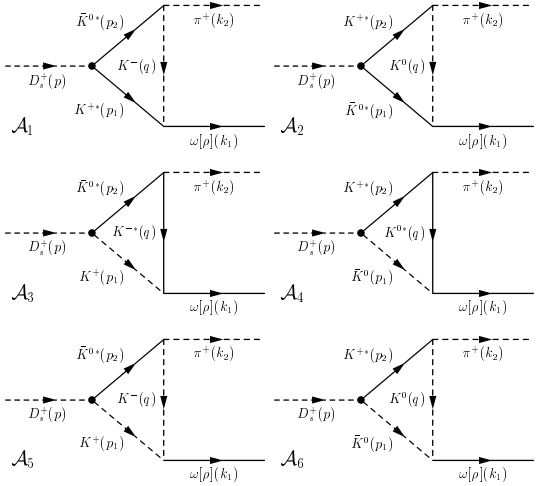}\\[-8mm]
\caption{\footnotesize{The $K, K^*$ meson contributions in the  
$D_s^+ \to \omega \pi^+$ and $D_s^+ \to \rho^0 \pi^+$  
decay amplitudes.}}
\label{fig-diagrams} 
\end{center} 
\end{figure} 

\begin{figure} 
\begin{center} 
\includegraphics[width=6cm]{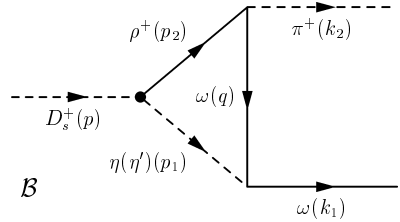}\\[-8mm] 
\caption{\footnotesize{ The intermediate $\eta$, $\eta^\prime, \rho^+$ contributions  
in the $D_s^+ \to \omega \pi^+$ decay.}}\label{fig-eta} 
\end{center} 
\end{figure} 

For the  weak current matrix elements between $D_s$ and vector or 
pseudoscalar final states we use a common form factor  decomposition 
\cite{PDG,FPPZ} with the form factors  
$F_+(q^2)$, $V(q^2)$, $A_{1,2} (q^2)$and $A_0(q^2)$.
For the $q^2$ dependence of the form factors 
we use results of \cite{FF}, based on a quark model calculation 
combined with a fit to lattice and experimental data. Ref.\cite{FF} 
provides a simple fit to their numerical results with the form factors 
$F_+(q^2)$, $V(q^2)$ and $A_0(q^2)$ described by double pole 
$q^2$ dependence 
\begin{equation} 
f(q^2)=\frac{f(0)}{(1-q^2/M^2)(1-\sigma q^2/M^2)}\;, 
\label{double} 
\end{equation} 
while single pole parameterization 
\begin{equation} 
f(q^2)=\frac{f(0)}{(1-\sigma q^2/M^2)}\;, 
\label{single} 
\end{equation} 
can be  used for $A_{1,2} (q^2)$, as the contributing resonances 
have masses farther away from the physical region (note that this parameterization 
applies also to $F_0$ form factor, which however does not contribute 
in the processes we discuss in this paper). The values of $f(0)$ and 
$\sigma$ are listed in Table \ref{formfaktorji1} and are taken from  \cite{FF}. 
We use $M=1.97$ $ \rm GeV$ in the expression for $A_0$, and  
$M=2.11$ $ \rm GeV$  for all the other   
form factors \cite{FF}.  Incidentally, the parameterizations of the form 
factors \eqref{double} and \eqref{single} 
 make all the loop diagrams in Figs. 
\ref{fig-diagrams} and \ref{fig-eta} finite. 

\begin{table} 
\begin{center} 
\begin{tabular}{c|cccccc} 
\hline 
 & $F_+$ & $V$  & $A_0$ & $A_1$ & $A_2$ & $F_{\eta_s,+}$ \\ \hline  
$f(0)$      & 0.72  & 1.04 & 0.67  & 0.57  & 0.42  &	0.78      
 \\ 
$\sigma$    & 0.2   & 0.24 & 0.2   & 0.29  & 0.58  &    0.23      \\ \hline 
\end{tabular} \\[3mm]
\label{formfaktorji1} 
\caption{\footnotesize{Form factors at $q^2=0$  \cite {FF}. The results in the first five  
columns are for $D_s \to K,K^* l \nu_l$ transitions.  
The last column stands  for the form factor appearing in  
the $D_s \to \eta_s l \nu_l$,  
(the $s \bar s$ component of $\eta, \eta'$)  
transition.}} 
\end{center} 
\end{table}
For the decay constants, defined through  
$\langle 0|\bar q\gamma^\mu \gamma_5 q|P(p)\rangle=if_P p^\mu $  
and 
$\langle 0|\bar q\gamma^\mu q|V(p)\rangle=g_V \varepsilon^\mu$, 
we use 
 $f_D=0.207\; \rm{GeV}$ and $f_{Ds}=1.13 f_D$ as obtained on the 
lattice \cite{fdlat}  
and for the rest  $f_K=0.16\;\rm{GeV}$, $|g{_K*}|=0.19\;\rm{GeV}^2$, $|g_\rho|=0.17\;\rm{GeV}^2$  
and $|g_\omega|=0.15\;\rm{GeV}^2$ coming from the experimental measurements \cite{PDG}. 
 
The amplitudes for the $D^+_s \to \omega \pi^+$ and $D^+_s \to \rho^0 \pi^+$ 
decays can be written as: 
\begin{align}  
{\cal A}(D^+_s \to  
\omega\pi^+)=\frac{G_F}{\sqrt{2}} &
\varepsilon 
\cdot k_2 \Big(\sum_i {\cal A}_i^{(\omega)}+{\cal B}\Big) \label{omega_amp} 
, \\ 
{\cal A}(D^+_s \to \rho^0\pi^+)=
\frac{G_F}{\sqrt{2}}&\varepsilon \cdot k_2 \sum_i {\cal A}_i^{(\rho)}\;\label{rho_amp} 
, 
\end{align} 
with $\varepsilon$  the helicity zero polarization vector of the 
$\omega$ or $\rho$ vector mesons, while  $k_2$ is the pion momentum.  
The reduced amplitudes ${\cal A}_i^{(\rho),(\omega)}$ and ${\cal B}$ correspond to the 
diagrams in Figs. \ref{fig-diagrams} and \ref{fig-eta} respectively. 
The explicit expressions  can be found in  Appendix of \cite{FPPZ}. The 
numerical values for  ${\cal A}_i^{(\rho),(\omega)}$ and ${\cal B}$ are given in Table \ref{results}. 
\begin{table}
\begin{center} 
\begin{tabular}{ccc} 
\hline  
$D_s^+ \to \omega \pi^+$& ${\cal A}_{iD}$ &${\cal A}_{iA}$\\
\hline
${\cal A}_1$ &$-0.7$ &$-0.7$\\
${\cal A}_2$ & $0.7$& $0.7$\\
${\cal A}_3$&$-1.1$&$3.3$\\
${\cal A}_4$ & $-1.4$&$1.5$\\
${\cal A}_5$ & $11.3$ &$-4.0$\\
${\cal A}_6$ & $12.5$&$-19.7$\\
${\cal B}_\eta$ & $1.3$& $-7.2$\\
${\cal B}_{\eta^\prime}	$ &$3.6$&$-3.7$\\ 
\hline 
\end{tabular}\\[3mm]
\caption{\footnotesize{ The dispersive ${\cal A}_{iD}$  and absorptive ${\cal A}_{iA}$  
parts of the amplitudes (in units of $10^{-3}$ ${\rm GeV}$)  for the  
$D_s^+ \to \omega \pi^+$ decay corresponding to the diagrams on 
Fig. \ref{fig-diagrams} (${\cal A}_i$) and Fig. \ref{fig-eta}  
(${\cal B}_{\eta,\eta'}$). 
 The 
amplitudes for the $D_s^+ \to \rho^0 \pi^+$ decay (neglecting the mass 
difference between $m_\rho$ and $m_\omega$) are obtained  by 
inverting the sign of ${\cal A}_{iD}, {\cal A}_{iA}$ for even $i$, 
while ${\cal B}_{\eta,\eta'}=0$.}} 
\label{results} 
\end{center} 
\end{table} 
Combining the above results we arrive at the  prediction   
\begin{equation} 
BR(D_s^+ \to \omega \pi^+)=3.0 
\times 10^{-3}. 
\end{equation} 
 Note that in this calculation we have used the 
factorization approximation for the diagram of Fig.  \ref{fig-eta}, 
which as stated above, does not work well for $D_s^+ \to \rho^+  
\eta,\eta'$ transition. Including hidden strangeness FSI   
to the $D_s^+\to \rho^+ \eta'$ 
decay mode   gives an order of 
magnitude smaller contribution.  
On the other hand one can use the experimental input to rescale 
the corresponding amplitudes. This results in the 
prediction 
$BR(D_s^+ \to \omega \pi^+)=4.4\times 10^{-3}$. 
We point out that the loop contributions are finite due to 
double pole parametrization of the form factors. 
If a single pole parametrization is used, one has to regularize 
the amplitudes. We found that the numerical results   do not 
change significantly in this case  when the cut-off scale is 
above but close enough to the $D_s$ meson mass. 
We can draw the conclusion that 
the experimental result for $BR(D_s^+ \to \omega \pi^+)$
can be understood  as a result 
of the combined effect of a spectator 
transition and FSI. Therefore, it makes the attempt to 
understand the $D_s^+ \to \omega \pi^+$ amplitude as a result 
of annihilation contributions unsuccessful. 

In the case of  the  $D_s^+ \to \rho^0 \pi^+$ transition, the FSI 
contributions alone 
result in  
\begin{equation} 
BR(D_s^+ \to \rho^0 \pi^+)_{\rm FSI}=0.7 \times 10^{-3} \,.\label{FSI_rho} 
\end{equation} 
This is almost exactly the same as our estimate of the upper bound 
on  the annihilation 
contribution \eqref{annihl}. Both contributions are equal or very close to   the present 
90\% CL upper bound. If there is no destructive interference 
between these two contributions and the contributions of FSI 
through higher resonances that we did not take into account,  
one hopes that the branching fraction for this decay  
will be determined in the near future. Our 
prediction is  in agreement with the results of other 
theoretical studies 
which give the rate for  
$D_s^+ \to \rho^0 \pi^+$ to be equal  
\cite{buccella2} or even larger than  
the rate for $D_s^+ \to \omega \pi^+$ decay \cite{hin,rosner}.  
 
However, one should consider possible 
cancellation that might occur. 
 Adding the FSI contribution and the maximal 
annihilation contributions \eqref{annihl} with alternating signs gives a fairly large 
interval  
\begin{equation} 
BR(D_s^+ \to \rho^0 \pi^+)=(0.05-3.5)\times 10^{-3}\,. 
\end{equation} 
We note that the experimental uncertainties reflected  in the input parameters can change the values 
for $BR(D_s^+ \to \rho^0 \pi^+)$ and $BR(D_s^+ \to \omega \pi^+)$ 
by about 20\%.  
 
Finally, we mention that the kind of FSI contributions we were 
considering in this paper is not the leading contribution in  the $D_s^+\to 
\phi \pi^+$ transition, which can proceed through 
spectator quark transition directly. Use of the factorization approximation for the weak vertex 
leads to a prediction $BR(D_s^+\to 
\phi \pi^+)=4.0\%$, which is already in excellent agreement with the 
experimental result of $3.6 \pm 0.9 \%$. 
We found that inclusion of FSI reduces the 
theoretical prediction from 4\% to  $\sim 3.6\%$.   The 
size of the shift also indicates 
that FSI of the type described in the present paper are  in the case of $D_s^+\to 
\phi \pi^+$ transition a second order effect. Note as well, that the 
size of the FSI correction is in agreement with the predictions for 
$BR(D_s^+\to \rho^0\pi^+)$ and $BR(D_s^+\to \omega \pi^+)$, which are 
an order of magnitude smaller than $BR(D_s^+\to\phi \pi^+)$. 
 
We summarize that the hidden strangeness final state 
interactions are very important in understanding the  
$D^+_s \to \omega \pi^+$ and $D^+_s \to \rho^0 \pi^+$ decay mechanism.  
The $D^+_s \to \omega \pi^+$ amplitude can be explained fully by this 
mechanism.  In the case of 
the $D^+_s \to \rho^0 \pi^+$ decay rate 
we obtain  a fairly large range due to possible cancellation 
between FSI and single pole contributions. 

The measurement of the   $D^+_s \to \rho^0 \pi^+$ decay rate 
will considerably improve  our understanding 
of the $D^+_s \to \rho^0 \pi^+$ decay mechanism. 
The hidden strangeness FSI might fully explain the observed decay rate
for  $BR(D_s^+\to \omega \pi^+)$. Finally, 
this kind of  FSI gives only subdominant 
contributions in the case of  $D_s\to \phi 
\pi, K K^*$ decays, which are well described by the factorization approximation. 

\vskip5mm 
\noindent 
{\bf {\large Acknowledgments}}\\[2mm] 
S.F., A.P. and J.Z. are 
supported in part by the Ministry of Education, Science and Sport of 
the Republic of Slovenia.

\end{document}